\documentclass[sn-mathphys,Numbered]{sn-jnl}

\usepackage{graphicx}%
\usepackage{multirow}%
\usepackage{amsmath,amssymb,amsfonts}%
\usepackage{amsthm}%
\usepackage{mathrsfs}%
\usepackage[title]{appendix}%
\usepackage{xcolor}%
\usepackage{textcomp}%
\usepackage{manyfoot}%
\usepackage{booktabs}%
\usepackage{algorithm}%
\usepackage{algorithmicx}%
\usepackage{algpseudocode}%
\usepackage{listings}%
\usepackage{docmute}

\theoremstyle{thmstyleone}%
\theoremstyle{thmstyletwo}%

\theoremstyle{thmstylethree}%

\usepackage{pdfpages}

\begin{document}


\title [mode = title]{TSUBF-Net: Trans-Spatial UNet-like Network with Bi-direction Fusion for Segmentation of Adenoid Hypertrophy in CT}             


\author[1,2]{Rulin Zhou}\email{2020111044@email.szu.edu.cn}
\equalcont{These authors contributed equally to this work.}

\author[3]{\fnm{Yingjie} \sur{Feng}}\email{2021280180@email.szu.edu.cn}
\equalcont{These authors contributed equally to this work.}

\author[4]{\fnm{Guankun} \sur{Wang}}\email{gkwang@link.cuhk.edu.hk}

\author[2]{Xiaopin Zhong}\email{xzhong@szu.edu.cn}

\author[2]{Zongze Wu}\email{zzwu@szu.edu.cn}

\author*[1, 5]{Qiang Wu}\email{sdzywq@szu.edu.cn}

\author[2,5,6]{Xi Zhang}\email{zh0005xi@szu.edu.cn}

\affil*[1]{\orgname{Department of Otorhinolaryngology Head and Neck Surgery, Shenzhen University General Hospital},  \city{Shenzhen}, \country{China}}

\affil[2]{\orgname{College of Mechatronics and Engineering, Shenzhen University},  \city{Shenzhen}, \country{China}}

\affil[3]{\orgname{College of Electronics and Information Engineering, Shenzhen University},  \city{Shenzhen}, \country{China}}

\affil[4]{\orgname{Department of Electronic Engineering, The Chinese University of Hong Kong},  \city{Hong Kong}, \country{China}}

\affil[5]{\orgname{Research Center of Medical Plasma Technology, Shenzhen University},  \city{Shenzhen}, \country{China}}

\affil[6]{\orgname{Shenzhen Lihuy Medical Tech. co. Ltd.},  \city{Shenzhen}, \country{China}}


\abstract{{Adenoid hypertrophy stands as a common cause of obstructive sleep apnea-hypopnea syndrome in children. It is characterized by snoring, nasal congestion, and growth disorders. Computed Tomography (CT) emerges as a pivotal medical imaging modality, utilizing X-rays and advanced computational techniques to generate detailed cross-sectional images. Within the realm of pediatric airway assessments, CT imaging provides an insightful perspective on the shape and volume of enlarged adenoids. Despite the advances of deep learning methods for medical imaging analysis, there remains an emptiness in the segmentation of adenoid hypertrophy in CT scans. To address this research gap, we introduce TSUBF-Net (\textbf{T}rans-\textbf{S}patial \textbf{U}Net-like Network based on \textbf{B}i-direction \textbf{F}usion), a 3D medical image segmentation framework. TSUBF-Net is engineered to effectively discern intricate 3D spatial interlayer features in CT scans and enhance the extraction of boundary-blurring features. Notably, we propose two innovative modules within the U-shaped network architecture: the \textbf{T}rans-\textbf{S}patial \textbf{P}erception module (TSP) and the \textbf{B}i-directional \textbf{S}ampling \textbf{C}ollaborated \textbf{F}usion module (BSCF). These two modules are in charge of operating during the sampling process and strategically fusing down-sampled and up-sampled features, respectively. Furthermore, we introduce the Sobel loss term, which optimizes the smoothness of the segmentation results and enhances model accuracy. Extensive 3D segmentation experiments are conducted on several datasets. TSUBF-Net is superior to the state-of-the-art methods with the lowest HD95: 7.03, IoU:85.63, and DSC: 92.26 on our own AHSD dataset. The results in the other two public datasets also demonstrate that our methods can robustly and effectively address the challenges of 3D segmentation in CT scans.}}




\maketitle
\section{Introduction}\label{sec1}

The main cause of upper airway obstruction in children and adolescents is adenoid hypertrophy (AH)~\cite{PEREIRA2018101}. AH typically manifests with symptoms such as snoring, nasal congestion, rhinorrhea, sinusitis, and hearing impairment, potentially leading to the characteristic "adenoid face" over prolonged durations~\cite{peltomaki2007effect}. Moreover, obstructive sleep apnea hypoventilation syndrome (OSAS) is a prevalent respiratory disorder during childhood sleep, primarily rooted in hypertrophy of the adenoids and tonsils~\cite{lu2022polysomnography}. If left unaddressed, this condition carries the risk of severe sequelae, encompassing neurocognitive aberrations, cardiovascular complications, and disruptions in growth patterns~\cite{scadding2010non}.

The principal detection modalities of AH encompass ultrasonography, X-ray, CT, MRI, and nasal endoscopy, among which CT examination holds notable significance owing to its commendable density resolution. CT scans excel in delineating tissue structures with macroscopic pathological anatomy. Particularly, CT transverse images provide a panoramic depiction of adenoids and the nasopharyngeal cavity, enabling accurate measurements of nasopharyngeal airway obstruction and concurrent observation of surrounding tissues~\cite{major2014agreement}. Consequently, employing CT for the segmentation of adenoid hypertrophy proves to be of paramount importance.

\begin{figure*}
    \includegraphics[width=1.27\textwidth, trim=0 0 0 0]{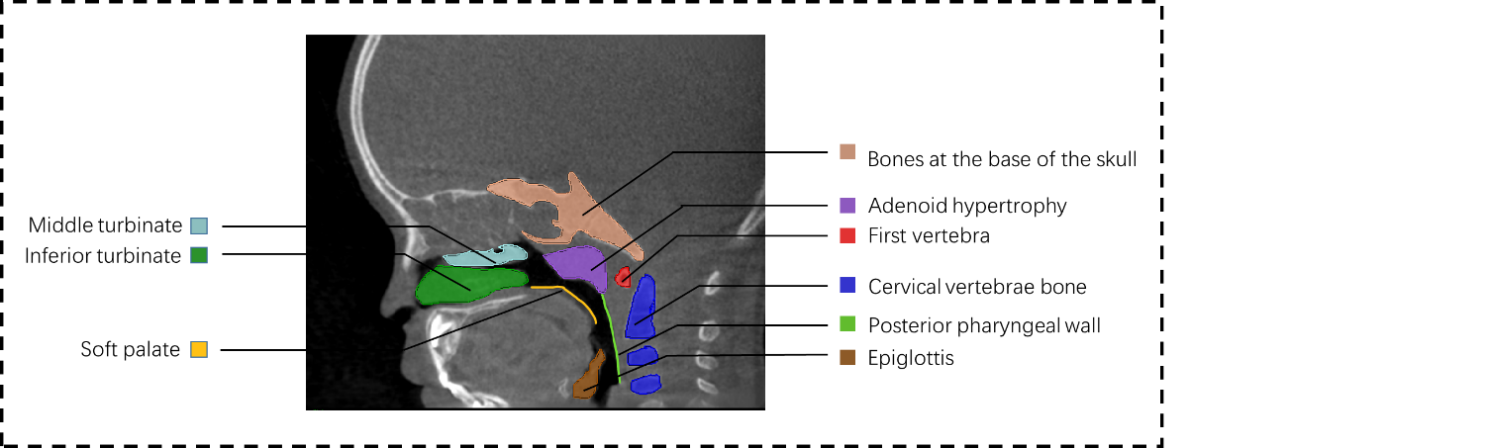}
    \caption{Schematic representation of tissue structures in CT of the head.}
    \label{fig:example}
\end{figure*}

Subsequently, through CT images, a broader physiological context is accessible, as illustrated in Fig.~\ref{fig:example}, encompassing structures such as the skull, vertebrae, soft palate, epiglottis, and turbinates. The specific target for segmentation is denoted in purple, representing the hypertrophied segment of the adenoids. CT reveals that the hypertrophied portion is located within a confined airway space, surrounded by numerous bones, muscles, and fascial tissues, necessitating careful consideration during surgical interventions. Surgeons traditionally rely on nasal endoscopy~\cite{shin2003pediatric} and oral endoscopy~\cite{yanagisawa1997endoscopic} for visualizing this intricate anatomy. However, nasal and oral endoscopy provide only a two-dimensional perspective with a limited field of view, presenting challenges during surgery. Consequently, physicians increasingly seek computer-assisted pre-segmentation and planning for adenoid surgery to alleviate the procedural burden. Consequently, the application of deep learning for analyzing adenoid hypertrophy in the nasopharynx is gaining popularity. Notably, there is currently a gap in the literature concerning deep learning-based segmentation specifically tailored for adenoid hypertrophy in CT scans.

In the realm of clinical surgery, the utilization of low-temperature plasma radiofrequency ablation has emerged as a pivotal intervention for addressing pharyngeal diseases, particularly in the surgical management of obstructive upper pharyngeal airway conditions, including Obstructive Sleep Apnea-Hypopnea Syndrome (OSAHS), attaining commendable efficacy. Nevertheless, inherent constraints exist in this procedure. Notably, plasma pharyngeal dilatation is often a multiplanar, one-stage undertaking, involving simultaneous intervention on multiple stenosis planes without a standardized protocol. Surgeons predominantly rely on experiential judgment, lacking quantitative preoperative planning methods, which may result in imprecise resection and recurrence. In this condition, leveraging adenoid segmentation results from the patient's CT scans for deep-learning-assisted preoperative planning is advantageous. Furthermore, we posit that CT scans offer a more comprehensive perspective compared to X-ray or endoscopy images. The high-resolution 3D images rendered by CT scans, depicting adenoid hypertrophy and its surrounding tissues, provide a nuanced portrayal of critical details such as the location, size, and shape of the hypertrophied segment. This comprehensive visualization allows for the anticipation of potential compression effects on adjacent structures, enabling the formulation of a judicious and precise surgical plan. Consequently, surgeons can engage in comprehensive preoperative planning, facilitating enhanced intraoperative efficacy.

\begin{figure*}
    \includegraphics[width=1\textwidth, trim=0 0 0 0]{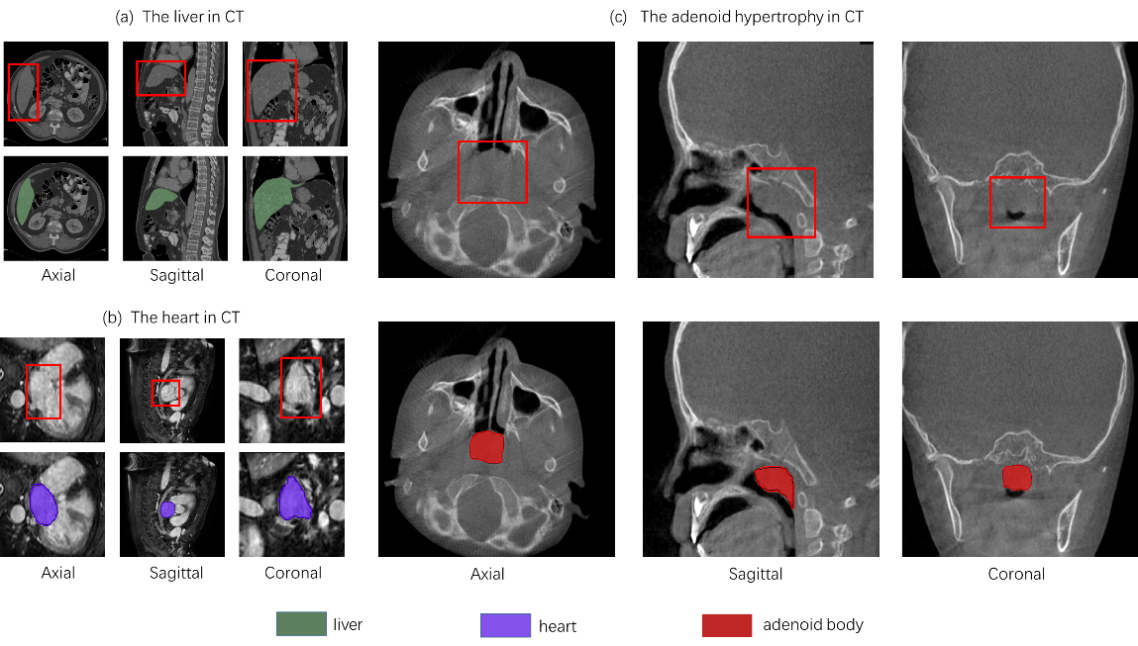}
    \caption{Common segmentation task vs. adenoid hypertrophy segmentation task: Unlike common segmentation tasks, the boundaries of the segmented objects in the adenoid hypertrophy segmentation task are vague and unclear.}
    \label{fig:example_2}
\end{figure*}

Nonetheless, the segmentation of adenoid hypertrophy encounters challenges that the task deviates from conventional medical 3D segmentation endeavors due to the relatively indistinct nature of its segmentation boundaries. As depicted in Fig.~\ref{fig:example_2}, typical segmentation tasks, such as delineating the liver and heart in CT scans, exhibit well-defined boundaries and precise localization. In contrast, the adenoid hypertrophic region in the Axial plane of CT scans presents challenges with less conspicuous segmentation boundaries. The anterior boundary aligns with the posterior edge of the nostril, while the posterior boundary and the delineation of lateral boundaries exhibit ambiguity. In the clinical surgical context, it is imperative to restrict the depth of the posterior border to mitigate the risk of inadvertent damage to the posterior vertebrae, prevertebral muscles, and fascia during surgical ablation. This aspect holds significant clinical relevance to minimize postoperative complications and promote optimal patient outcomes.

Therefore, we have formulated the Trans-Spatial UNet-like Network with Bi-direction Fusion (TSUBF-Net), a 3D medical image segmentation network grounded on a U-shaped architecture, tailored specifically for the segmentation task of adenoid hypertrophy in CT scans. To solve the challenge of weak boundedness in the adenoid dataset, our network prioritizes spatial features and enhances model spatial perception through the integration of the Trans-Spatial Perception Module (TSP). Besides, we design the Bi-direction Sample Collaborated Fusion Module (BSCF) within both down-sampling and up-sampling processes to incorporate the feature fusion process of the U-shaped network. Additionally, to optimize the smoothness of the segmentation targets, adjustments have been made to the loss function. Specifically, we employ the gradient operator to compute gradients in the x, y, and z directions of the 3D CT data, incorporating these computed values as components of the loss function. Through extensive experiments, our proposed methods present superior performance and robustness in the segmentation of adenoid hypertrophy. {In summary, our key contributions are as follows:
\begin{itemize}
    \item[--] We propose a 3D segmentation framework TSUBF-Net within the U-shaped network architecture, which provides valuable pioneering explorations and experiences for the task of 3D segmentation of adenoid hypertrophy. 
    \item[--] We design the Trans-Spatial Perception Module (TSP) which emphasizes the spatial features through a layer-based attention mechanism across dimensions to improve 3D segmentation models in spatial capabilities.
    \item[--] We adopt the Bi-direction Sample Collaborated Fusion Module (BSCF) for the fusion of up-sampled and down-sampled features in U-shaped networks, which improves the segmentation accuracy of the model.
    \item[--] The Sobel loss term is integrated with the loss function to improve the segmentation accuracy while improving the smoothness of the segmentation results.
    \item[--] According to experiments, our proposed framework can effectively handle the segmentation challenge of fuzzy boundaries associated with adenoid hypertrophy. In addition, we also introduced other image types of organ segmentation data for experiments, such as MRI data, and the experimental results show that our method still shows excellent performance on other 3D organ segmentation data sets, and has the potential to be widely applied.
\end{itemize}
}

\section{Related Work}\label{sec2}
\subsection{Deep Learning Segmentation to Adenoid Hypertrophy}\label{subsec2}

With the development of deep learning, the application in medical imaging is becoming an emerging area. More and more methods are proposed to target the disease problem of adenoid hypertrophy. Wang \textit{et al.}~\cite{wang2020evaluation} used ultrasound to evaluate adenoid hypertrophy. Zhao \textit{et al.}~\cite{zhao2021automated} designed a CNN-based framework to train on clinical data of avatars. Shen \textit{et al.}~\cite{shen2020deep} classified adenoid hypertrophy by locating key points and devising a novel regularized terminology based on patient's X-rays. Liang \textit{et al.}~\cite{liang2022efficient} used a convolutional neural network to analyze children's behavior at night during the night to classify normal snoring and abnormal snoring. Zheng \textit{et al.}~\cite{zheng2022contrastive} used a novel multiscale hierarchical network MIB-Anet to classify adenoid images from nasal endoscopy in four classes. Alshbishiri \textit{et al.}~\cite{alshbishiri2021adenoid} sampled a 2D-UNet network to segment the adenoids. Wang \textit{et al.}~\cite{ronneberger2015u} used a 2D-UNet network to classify the adenoid hypertrophic portion of the adenoids from X-rays. Despite numerous studies on the segmentation of 2D adenoid hypertrophy, there is little work on the 3D segmentation of adenoid CT data.

\subsection{CNN-Based Segmentation Networks}
Since the introduction of U-Net~\cite{ronneberger2015u}, CNN-based networks have achieved state-of-the-art results in various 2D and 3D medical image segmentation tasks~\cite{wang2023domain}~\cite{wang2023rethinking}~\cite{huang2020unet}~\cite{zhou2018unet++}. 
In the case of 3D medical image segmentation, the complete volume is usually processed into a sequence of 2D slices, which are stacked after the 2D data has been segmented. Çiçek \textit{et al.}~\cite{cciccek20163d} used the architecture of U-Net for the task of volume segmentation using a 3D convolutional kernel for feature extraction. licensee \textit{et al.}~\cite{isensee2021nnu} proposed an adaptive framework nnUNet based on both 2D and 3D Adaptive framework which adapts the segmented target for feature extraction at multiple scales. Roth \textit{et al.}~\cite{roth2017hierarchical} proposed a 3D fully convolutional segmentation network mainly for multi-organ segmentation. Schlemper \textit{et al.}~\cite{schlemper2019attention} proposed a novel Attention Gate (AG) model that automatically learns to attend to different shapes and sizes of target structures. The mu-Net proposed by~\cite{seo2019modified} adds a residual path with de-convolution and activation operations at the skip junction of up-sampling and down-sampling to extract high-level global features for small object inputs and high-level features for large object inputs with high-resolution edge information, and it performs well for 3D segmentation of the liver in CT.

Merged U-net~\cite{xie2022merged} deals with the task of 2D segmentation of bone tumor radiographs. The noteworthy part of this model is merge gates are used to deal with the problem of blurred segmentation boundaries. Among other things, the merge gate is used in the process of connecting the down-sampled features to the up-sampled features of the model, making the feature connections selective. {Other than that, as a deep learning framework for image semantic segmentation, DeepLab v3+~\cite{chen2018encoder} uses dilated convolution to obtain a larger receptive field and performs better in the precision processing of segmentation boundaries. Kubilay Muhammed Sunnetci \textit{et al.}~\cite{SUNNETCI2024157} proposed a new segmentation model architecture based on DeepLab v3+ for MRI images of Comprehensive Parotid Gland Tumor disease, and has better processing capabilities for tumor edge information.}

\subsection{Visual Transformer-based Segmentation Networks} 
Visual transformer (VIT) has become popular recently due to its excellent performance in deep learning segmentation. VIT was originally proposed for the NLP domain~\cite{dosovitskiy2020image}, then Bichen \textit{et al.}~\cite{wu2020visual} used Transformer for visual processing problems, which is becoming more and more widely used in computer vision~\cite{child2019generating}~\cite{kitaev2020reformer}. The core idea of VIT is the self-attention mechanism, which can extract more global information and better connect the spatial information. Karimi \textit{et al.}~\cite{karimi2021convolution} proposed to divide the volume image into three-dimensional modules, and using a module with an adjacency-attention mechanism gives better results than the convolution module in some work. Hu \textit{et al.}~\cite{cao2022swin} proposed a model that uses a pure Transformer to replace the convolution operation of U-net to perform local-convolutional operations. {Qiu \textit{et al.}~\cite{qiu2024agileformerspatiallyagiletransformer} propose to introduce spatial dynamic components into ViT-UNet to efficiently capture the features of target objects with different appearances.}

\subsection{Hybrid segmentation methods} 
In addition to pure CNN or ViT-based segmentation network design, more and more work has begun to explore hybrid frameworks. Wang \textit{et al.}~\cite{wang2021transbts} applied the ViT to segmentation of MRI brain tumors in 3D. Xie \textit{et al.}~\cite{xie2021cotr} proposed a new framework, CoTr, that efficiently connects the CNN and the ViT to achieve accurate 3D medical image segmentation. Hatamizadeh \textit{et al.}~\cite{hatamizadeh2022unetr} designed a model of UNETR which completely uses the Transformer as an encoder to down-sample the data and the CNN as a decoder to up-sample and get more global information.  {Swin-UNETR~\cite{hatamizadeh2021swin} uses Swim-Transformer to compute the self-attention by utilizing a shift window, which is connected to the FCNN-based decoder for feature extraction. It can effectively extract the feature of the segmented task, but in the processing of each layer separately, it does not get the edge information well, resulting in the segmentation result is not ideal. nnFormer model proposed by Zhou \textit{et al.}~\cite{zhou2021nnformer} combines convolutional and self-attention mechanisms interleaved organically, replacing the single CNN or Transformer module. However, in the task of 3D image segmentation, the performance is not good to obtain the segmentation results with smooth edges, but from a certain layer, the comparison and label overlap. Shaker \textit{et al.}~\cite{shaker2024unetr} devised a new Efficient Attention Module (EPA), which uses a pair of spatial and channel-based attention mechanism modules with interdependent shared weights to efficiently learn discriminative features in both spatial and channel directions.}

In the medical 3D image segmentation task, the models with outstanding results are all hybrid models of convolutional U-net and Transformer. As the current SOTA model, it is necessary to have a deeper understanding of the internal structure of the UNETR++~\cite{shaker2024unetr}. The biggest contribution of UNETR++ is the design of an EPA block, which allows UNETR++ to obtain good results with the least number of parameters and computational cost. One of the main enhancements of the EPA block is to artificially set up a two-headed self-attention mechanism for the space and the channel and to use the same Q and K matrices thus emphasizing the strong correlation between the two. {However, we found that UNETR++ is not very effective in dealing with the problem that the target boundary is not very clear. There are even cases where the segmentation surface is not uniform sometimes. In this regard, our conjecture reason is that the spatial attention in the EPA block in UNETR++ has the problem of information loss during the linear dimensionality reduction process. However, UNETR++ also works well on common organ segmentation surfaces. The explanation given for this contradiction is that the corresponding missing part of the information is supplemented when the down-sampled features are connected to the up-sampled ones. This may also explain the fact that UNETR++ is abnormal for a few slices when organs are delineated by thin slats.}

\section{Methodology}

We introduce a Trans-Spatial UNet-like Network with Bi-direction Fusion(TSUBF-Net) to enhance the precision of the 3D segmentation model concerning the challenge of adenoid hypertrophy. As expounded in the above section, our analysis reveals that the adenoid hypertrophy task exists in the absence of discernible geometric boundaries within the image. In light of this intrinsic trait, we have devised two modules: the Trans-Spatial Perception (TSP) module and the Bi-direction Sample Collaborated Fusion (BSCF) Module. Moreover, to augment the overall smoothness of the segmentation output, we introduce an explicit measure of segmentation smoothness, which bases on the application of the 3D Sobel operator and is integrated in the loss function.

\subsection{Network framework}

\begin{figure*}
    \centering
    \includegraphics[width=1\textwidth]{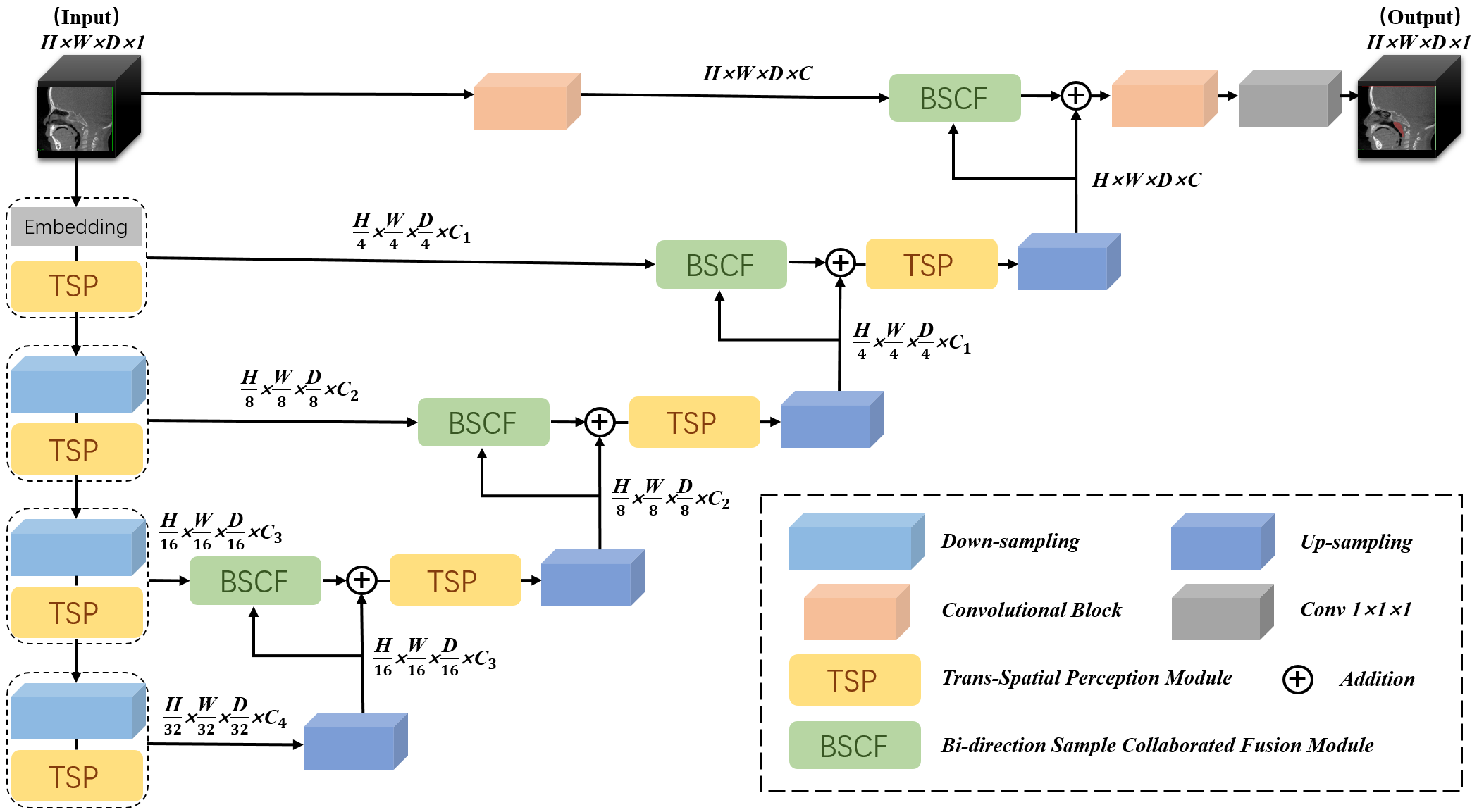}
    \caption{Overview of the proposed TSUBF-Net framework. TSUBF-Net adopts a U-shaped structure. Among them, the design of combining convolutional feature extraction, TSP module, and BSCF module is adopted in both up-sampling and down-sampling paths. In the model down-sampling path, the original image first passes through the patch embedding layer, and the feature extraction is performed on the original data through the operation of convolution, which makes the feature size become H/4*W/4*D/4*$C_1$. Similarly, the feature size of the 3D will be decreasing exponentially with the four times of the down-sampling structure, and its channel features will be increasing with the down-sampling.}
    \label{fig:main2}
\end{figure*}

We formulate TSUBF-Net, a framework that integrates the TSP module and BSCF module to facilitate the segmentation of adenoid hypertrophy. An overview of this architecture is shown in Fig.~\ref{fig:main2}.

Due to the advantages of multi-scale information fusion, our proposed TSUBF-Net follows a U-shaped architectural configuration. Notably, to combine the advantages of convolution and attention, the integration of convolutional feature extraction and TSP module is adopted in both up-sampling and down-sampling paths. In the down-sampling phase of the model, the initial image passes through the patch embedding layer, subsequently undergoing convolutional feature extraction, resulting in a down-sampled image resolution, reduced to one-quarter of the original size. Concomitantly, the dimension of the 3D feature representation undergoes continuous and exponential diminishment throughout four successive down-sampling stages, while its channel features increase.

Furthermore, we introduce a BSCF module within the interconnecting pathway between the down-sampling and up-sampling stages, facilitating the fusion of feature information from both these phases. Following a sequence of four up-sampling operations, the feature dimensions are fully restored to the original size.

\begin{figure*}
    \centering
    \includegraphics[width=1\textwidth]{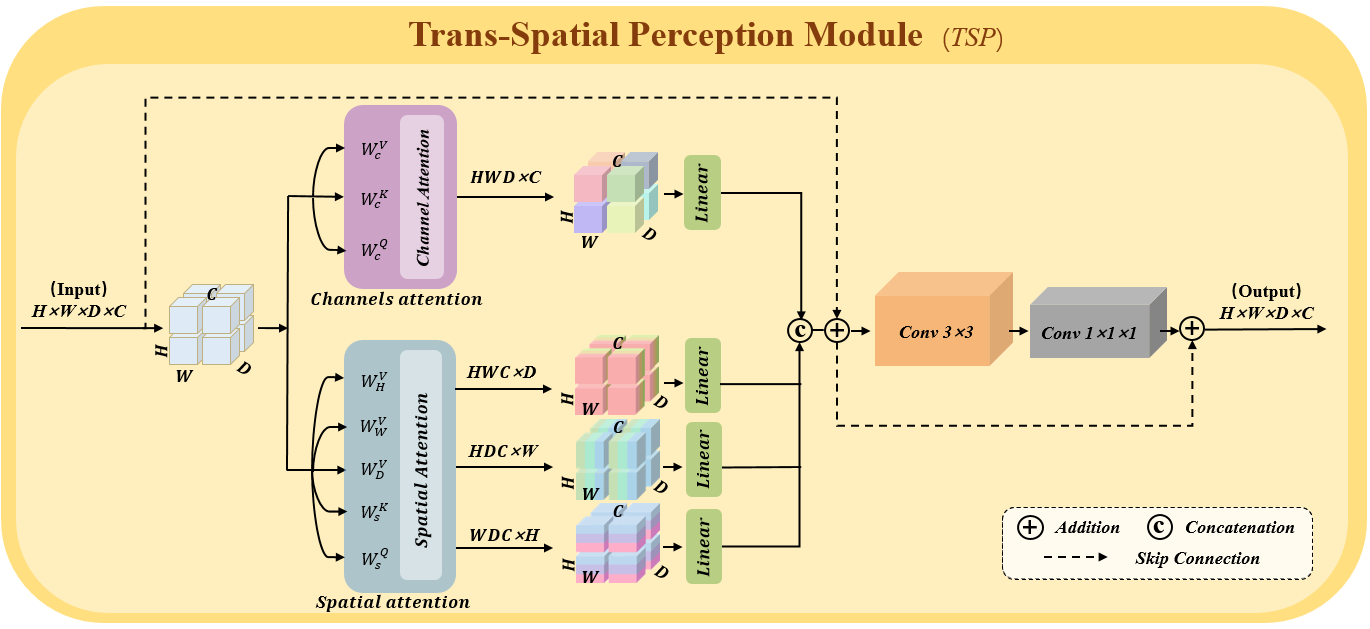}
    \caption{Detailed framework of the TSP module: the proposed TSP module is divided into a channel attention module and an inter-layer attention module, in which the channel attention carries out one head of attention in the channel dimension; the inter-layer attention consists of inter-layer attentions in three directions so that the inter-layer attention module is a three-head of attention, where the Q and K matrices are shared. Then it goes through a series of convolutional layers.}
    \label{fig:tsp}
\end{figure*}

\subsection{Trans-Spatial Perception Module}

To address the inherent absence of distinct geometric boundaries in the adenoid hypertrophy segmentation, we recognize the necessity for augmenting our attention module with spatial perception capabilities. Therefore, we introduce the TSP module. It is designed to enable effective global attention mechanisms, thereby adeptly capturing intricate spatial and channel feature representations. Given the 3D image segmentation features encompass four dimensions, we extend attention operations to each of these dimensions. As a result, the TSP module is predominantly comprised of a four-head attention mechanism, as depicted in Fig.~\ref{fig:tsp}. We categorize the four-head attention into two distinct components: channel attention and spatial attention. The latter emphasizes the spatial features interplay between different layers, with three heads dedicated to inter-layer attention across the height, width, and depth dimensions of the 3D image.

We define the input of the TSP module as $\boldsymbol{x} \in \mathbb{R}^{H \times W \times D \times C}$, then initiate a linear mapping process to derive the distinct Q, K, and V matrices essential for the attention mechanism. Specifically, this entails the generation of a total of 8 mapping matrices. The three constituent sub-attentions within the spatial attention encompass height attention, width attention, and depth attention. The weights assigned to the linear layers responsible for Q and K operations within the spatial attention are shared across all three layers of attention. Therefore, the spatial attention mechanism is computed concomitantly with the channel attention:

\begin{equation}
\begin{array}{c}
\widehat{X_{L}}=LA\left(Q_{\text{s}}, K_{\text{s}}, V_{\text{h}}, V_{\text{w}}, V_{\text{d}}\right) \\
\widehat{X_{C}}=CA\left(Q_{\text{c}}, K_{\text{c}}, V_{\text{c}}\right)
\end{array}
\end{equation}

where $\widehat{X_{L}}$ and $\widehat{X_{C}}$ denote the spatial attention map and the channel attention map, respectively. $LA$ is the spatial attention module, while $CA$ is the channel attention module. There are three more sub-modules in $LA$.$Q_{\text {s}}$, $K_{\text {s}}$, $V_{\text {h}}$, $V_{\text {w}}$, $ V_{\text {d}}$, $Q_{\text {c}}$, $K_{\text {c}}$, $V_{\text {c}}$ are spatial shared queries, spatial shared keys, height value layer, width value layer, depth value layer, channel queries, channel keys, channel value layer, respectively.

The attention mechanism necessitates three primary matrices: the query matrix, the key matrix, and the value matrix, all with the same shape. The computation of this attention is executed through the multiplication of the query matrix with the transposed key matrix, followed by a subsequent application of the softmax function to gauge the degree of similarity between each feature and its counterparts. These computed similarities are then applied to weight the corresponding projections within the value matrix, culminating in the derivation of the ultimate layer-specific attention map. {The meaning of this attention calculation formula is roughly the same as that of the attention calculation mechanism proposed by Bahdanau \textit{et al.}~\cite{Bahdanau_2014}, and finally our self-attention mechanism is defined as follows:}

\begin{equation}
\mathrm{\mathcal{A}}=\operatorname{Softmax}\left(\frac{Q_{\text {s}}, K_{\text {}}^{T}}{\sqrt{d}}\right) \cdot V
\end{equation}

Where, $Q$ denotes the query matrix, $K$ denotes the key matrix, $V$ denotes the value matrix, and $\boldsymbol{d}$ is the size of each vector.

\textbf{Spatial Attention:} The spatial attention module encompasses three distinct layer-specific attention modules. It is essential to reshape the feature values' dimensions before the processing of different attentions. More precisely, for height attention, we reshape $Q_{\text {s}}$, $K_{\text {s}}$, and $V_{\text {h}}$ with dimensions of $\mathbb{R}^{H \times W \times D \times C}$ to $\mathbb{R}^{WDC \times H}$. Similarly, for width and depth attention, we transform $Q_{\text {s}}$, $K_{\text {s}}$, and $V_{\text {w}}$ to dimensions of $\mathbb{R}^{HDC \times W}$ and $\mathbb{R}^{HWC \times D}$. Subsequently, the attention operation is performed independently across these three reshaped dimensions.

The attention output is further processed through a linear layer, which facilitates the extraction of features with reduced dimensionality, specifically shrinking it to one-quarter of its original size. This condensed representation is then arranged in a stacked fashion. In more precise terms, we obtain $\widehat{X_{L}} \in \mathbb{R}^{H \times W \times D \times \tfrac{3}{4} C}$ following the application of spatial attention module.

\textbf{Channel Attention: }It is leveraged to acquire channel information through the calculation of $Q_{\text{c}}$, $K_{\text{c}}$, and $V_{\text{c}}$ features, which share the same shape $\mathbb{R}^{HWD \times C}$. {It is imperative that the shape of the fused features align with the input, necessitating a dimension reduction via a linear layer to extract the hidden size and compress it to $1/4$ of its original dimension. }Consequently, we obtain $\widehat{X_{C}} \in \mathbb{R}^{H \times W \times D \times \frac{C}{4}}$ following the execution of the channel attention mechanism.

Concludingly, we execute a summative fusion process, where the outputs stemming from both attention modules transform a convolutional block, thereby yielding an enriched feature representation. The ultimate output of the TSP module, denoted as $\widehat{X}$, is characterized by the following form:

\begin{equation}
    \widehat{X}=Conv_{1}\left(Conv_{3}\left(Concat\left(\widehat{X_{L}}, \widehat{X_{C}}\right)\right)\right)
\end{equation}

Where $\widehat{X_{L}}$ and $\widehat{X_{C}}$ are spatial and channel attention, respectively. $Conv_{1}$ and $Conv_{3}$ are 1×1×1 and 3×3×3 convolutional blocks, respectively.

\begin{figure*}
    \centering
    \includegraphics[width=1\textwidth]{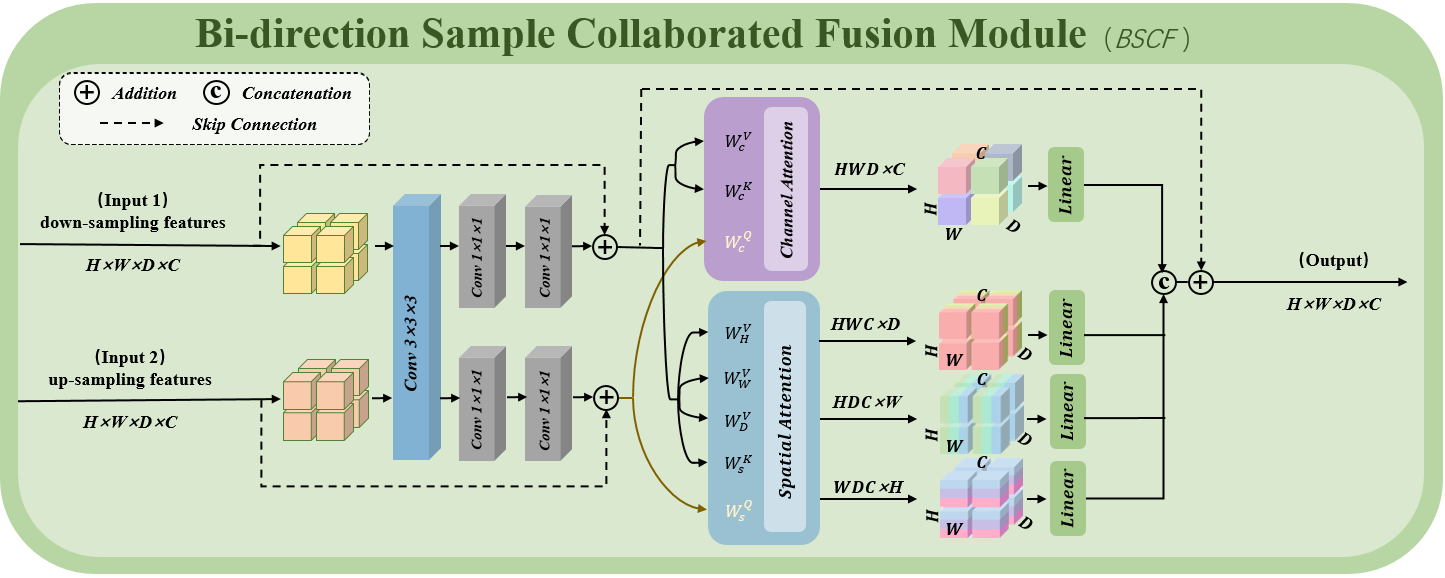}
    \caption{Detail framework of the BSCF module: the proposed BSCF module utilizes the same 3×3×3 convolution kernel for the first step of processing the up and down sampled features, thus emphasizing the consistency between the two, immediately followed by 1×1×1 convolution for further feature extraction, and then finally the attention mechanism is used to achieve the correction of the up-sampled feature information based on the up-sampled segmentation information. To emphasize the spatial features, TSP is used for the attention mechanism here.}
    \label{fig:bscf}
\end{figure*}

\subsection{Bi-direction Sample Collaborated Fusion Module}

In the TSBFU-Net framework, a combination of up-sampling and down-sampling processes is employed. The features acquired through the down-sampling phase encompass the geometric contour information inherent to the original image, while those obtained in the up-sampling phase contain more relevant information about the segmentation target. Given the peculiar attribute of the adenoid hypertrophy segmentation task, characterized by the absence of clear geometric boundaries, it becomes paramount to obtain a wealth of segmentation-relevant features. Consequently, our framework orchestrates the fusion of down-sampled geometric contour features with up-sampled segmentation features. In contrast to the direct connections in conventional U-shaped networks, we design the Bi-direction Sample Collaborated Fusion Module for intricate fusion inspired by Merged U-Net~\cite{xie2022merged}.

The BSCF module, as illustrated in Fig.~\ref{fig:bscf}, applies a uniform 3×3×3 convolution kernel in the initial processing step for both up-sampled and down-sampled features. This shared convolutional operation is designed to improve the congruence and alignment between these two sets of features. Subsequently, a 1×1×1 convolutional operation is introduced to facilitate further feature extraction. In this process, the attention mechanism effectively rectifies the information of up-sampled features based on segmentation-related cues derived from the upper sampling phase.

Based on the TSP module, the BSCF module receives the up-sampled feature $X_{u}\in \mathbb{R}^{H \times W \times D \times C}$ and the down-sampled feature $X_{d}\in \mathbb{R}^{H \times W \times D \times C}$ both sharing the identical dimensions of $H\times W\times D\times C$. As elucidated in Fig.~\ref{fig:bscf}, a series of convolutional operations is initially conducted to obtain the depth features $\widehat{X_{u}}\in \mathbb{R}^{H \times W \times D \times C}$ and $\widehat{X_{d}}$. The computational procedure is expounded as follows:

\begin{equation}
    \begin{array}{c}
         \widehat{X_{u}}=Conv1_{u}\left(Conv1_{u}\left(Conv3_{s}\left(X_{u}\right)\right)\right) \\
         \widehat{X_{d}}=Conv1_{d}\left(Conv1_{d}\left(Conv3_{s}\left(X_{d}\right)\right)\right)
    \end{array}
    \label{Eq:MA}
\end{equation}

Where, $Conv3_{s}$ is the 3×3×3 convolution kernel. $Conv1_{u}$ is the 1×1×1 convolution kernel for the up-sampled features. Note that the two $Conv1_{u}$ appearing in Eq~\ref{Eq:MA} are not the same. The inner $Conv1_{u}$ expands the original feature channel size to twice the original size, while the outer $Conv1_{u}$ reduces the feature channel length back to the original size. The design of $Conv1_{d}$ is the same. Subsequent to the feature extraction phase, the attention operation is executed. Diverging from the self-attention mechanism in the TSP module, in the BSCF module, the up-sampled features are designated as queries, while the down-sampled features assume the roles of keys and values. Q, K, and V matrices are derived by a linear mapping procedure, which will generate eight mapping matrices.

\subsection{Loss Function with Sobel Gradient}

In order to indistinct delineation of the segmentation target, we improve the loss function through the incorporation of the gradient operator. Specifically, we have elected to employ the Sobel gradient operator~\cite{gupta2013sobel} to formulate the Sobel loss so as to minimize the global gradient magnitude as much as possible.

The original loss function consists of two components including the soft dice loss and the cross-entropy loss~\cite{milletari2016v}, which is defined as follows:

\begin{equation}
    \mathcal{L}(\mathrm{Y}, \mathrm{P})=1-\sum_{i=1}^{l}\left(\frac{2 \times \sum_{v=1}^{V} Y_{v, i} \cdot P_{v, i}}{\sum_{v=1}^{V} Y_{v, i}^{2}+\sum_{v=1}^{V} P_{v, i}^{2}}+\sum_{v=1}^{V} Y_{v, i} \log P_{v, i}\right)
\end{equation}

Where $l$ denotes the number of classes; $V$ denotes the number of voxels; and $Y_{v,i}$ and $P_{v,i}$ denote the ground truth and output probability at voxel $v$ of class $i$, respectively.

The Sobel operator is a discrete differential gradient-based operator that is employed to compute a sparse approximation of image intensities for edge detection. In our approach, we adopt the Sobel operator with a (3,3,3) kernel to compute gradient maps along the X, Y, and Z axes directions for the segmentation result. These gradient maps are subsequently averaged to yield $G_{x}$, $G_{y}$, and $G_{z}$, respectively, based on the number of vectors. These gradients are then further subjected to a global gradient averaging procedure based on the Euclidean distances.

The conventional Sobel gradient operator typically employs a 3x3x3 three-dimensional matrix across all three dimensions. However, in our specific context, we need to account for the inherently three-dimensional characteristics of CT images. Therefore, our optimization objective extends beyond individual layers, emphasizing the holistic three-dimensional context. As a result, we have formulated the following Sobel operator vectors tailored to the x, y, and z dimensions:

\begin{equation}
    \begin{array}{c}
         H_{x}=\begin{bmatrix}
                \left[-1,0,1\right] & \left[-2,0,2\right] & \left[-1,0,1\right] 
             \\
                \left[-2,0,2\right] & \left[-4,0,4\right] & \left[-2,0,2\right]
            \\
                \left[-1,0,1\right] & \left[-2,0,2\right] & \left[-1,0,1\right] 
        \end{bmatrix} \\
        H_{y}=\begin{bmatrix}
                \left[-1,-2,-1\right] & \left[0,0,0\right] & \left[1,2,1\right] 
            \\   
                \left[-2,-4,-2\right] & \left[0,0,0\right] & \left[2,4,2\right]
            \\
                \left[-1,-2,-1\right] & \left[0,0,0\right] & \left[1,2,1\right]  
        \end{bmatrix} \\
        H_{z}=\begin{bmatrix}
                \left[1, 2, 1\right]     & \left[2, 4, 2\right]     & \left[1, 2, 1\right] 
            \\
                \left[0, 0, 0\right]     & \left[0, 0, 0\right]     & \left[0, 0, 0\right]
            \\
                \left[-1,-2,-1\right] & \left[-2,-4,-2\right] & \left[-1,-2,-1\right]  
        \end{bmatrix}
    \end{array}
\end{equation}

Considering our input image represented as $I\in \mathbb{R}^{M \times N \times P}$, the convolution operation, denoted as $\otimes$, yields the edge gradients in the three dimensions, which can be succinctly expressed as $G_{x}$, $G_{y}$, and $G_{z}$:

\begin{equation}
\begin{aligned}
G_{x}=\text{\textit{I}}\otimes{H_{x}}\\
G_{y}=\text{\textit{I}}\otimes{H_{y}}\\ 
G_{z}=\text{\textit{I}}\otimes{H_{z}}
\end{aligned}
\end{equation}

where $G_{x}$, $G_{y}$, and $G_{z}$ denote the mean value of the gradient of the segmentation result in the direction of X, Y, and Z axes, respectively. {We carefully studied the Sobel algorithm proposed by Canny {et al.}~\cite{Canny1987} for edge detection in images.} The final loss function for this part is defined as:

\begin{equation}
    \centering
    \mathcal{L_{\text{sobel}}}=\mathcal{\lambda}\sum_{i}^{l}\sqrt{G_{x}^{2}+G_{y}^{2}+G_{z}^{2}}
\end{equation}

where $\lambda$ denotes the weight, which is a hyper-parameter with the default value of 0.1. The final loss function consists of these two parts together:

\begin{equation}
    \centering
    \mathcal{L_{\text{total}}}=\mathcal{L}(\mathrm{Y}, \mathrm{P})+\mathcal{L_{\text{sobel}}}
\end{equation}

\section{Experiments}

\subsection{Datasets}

\textbf{Adenoid Hypertrophy Segmentation Dataset} is a comprehensive dataset that we present focused on adenoid hypertrophy. As shown in Table~\ref{tab:dataAHSD}.)This dataset was meticulously curated in collaboration with Shenzhen University General Hospital, Guangdong Province, Shenzhen, China, marking a significant innovation in the field. The data acquisition process underwent rigorous ethical scrutiny, receiving formal approval from the Ethical and Welfare Committee of Shenzhen University General Hospital (Approval No: KYLL-20230402A. Throughout our research, we maintained unwavering adherence to the ethical protocols and mandates outlined by the committee, ensuring the utmost protection of participants' rights and privacy. AHSD encompasses head CT images derived from a cohort of 240 patients, comprising 227 individuals diagnosed with adenoid hypertrophy and 13 individuals categorized as normal cases. {We collected this data using Imaging Sciences International Company specification model I-CAT FLX. All of our CT data volumes were 16D x 13H cm, and the number of slice layers along the Z axis was 533, each slice was 0.3 mm. We would like to be able to make this dataset public, but we are still going through the application and approval process due to the privacy of the data.} In our investigation, the primary focus was the precise segmentation of adenoid hypertrophic regions. Notably, from the pool of 227 patients exhibiting adenoid hypertrophy, we randomly designated 189 instances for training purposes, with the remaining 38 instances earmarked for the evaluation of our proposed methodology.

\begin{table}[!h]
\centering
\renewcommand{\arraystretch}{1.3} 
\caption{Detailed data information of AHSD}
\label{tab:dataAHSD}
\begin{tabular}{ccc}
        \hline
         \textbf{Whether the airway is abnormal}& \textbf{normal} & \textbf{unusual} \\
         \hline
         \textbf{Total quantity}& 13 & 227 \\
         \textbf{Sex (male/female)}& 6 / 7 & 106 / 121 \\
         \textbf{Age (mean±SD)}& 5.78±0.82 / 5±2.68 & 6.53±2.15 / 5.95±2.26 \\
         \hline
    \end{tabular}
\label{tab:my_label}
\end{table}

\textbf{Automatic Cardiac Diagnosis Challenge~\cite{bernard2018deep}} is introduced to demonstrate the effectiveness of our methods in a broader comparison. Designed for the purpose of automated cardiac diagnostics, ACDC encompasses cardiac MRI images acquired from a cohort of 100 patients during authentic clinical examinations. These images are meticulously annotated with segmentation information for key cardiac structures, including the right ventricle (RV), left ventricle (LV), and myocardium (MYO). The patient samples were thoughtfully selected to represent a diverse range of clinical scenarios, encompassing individuals with normal cardiac function, myocardial infarction, dilated cardiomyopathy, hypertrophic cardiomyopathy, and patients displaying abnormal right ventricle morphology. In consonance with the data splitting strategy akin to the Reformer framework~\cite{kitaev2020reformer}, the dataset was partitioned into three distinct subsets. Specifically, 70 images were designated for training, 10 samples were earmarked for validation, and the remaining 20 samples underwent rigorous evaluation in our experiments. {We chose this dataset for the following reasons: It is an MRI dataset, which can verify whether our algorithm has the same good performance in MRI 3D data as in CT; In addition, the ACDC data set has smooth boundaries, and there are no obvious boundaries in the original image, which is also conducive to evaluating the effectiveness of our model in processing edge smoothness on other data sets.}

{\textbf{MSD-Lung Dataset~\cite{Antonelli_2022}} is introduced to demonstrate the validity of our method on other publicly available CT datasets. The MSD-Lung Dataset is a subtask in the Medical Segmentation Decathlon (MSD). The objective of the dataset was to segment lung tumors, which consisted of preoperative thin-section CT scan images of 96 patients with non-small cell lung cancer. The data was acquired via the Cancer Imaging. Similar to the hypertrophied adenoids in AHSD that make up only a small portion of the head CT, lung tumors also make up a relatively small portion of the CT images. This dataset was chosen because of the similarities with AHSD, both of which segment small areas of the scene in a large field-of-view image. We believe that the two datasets are similar to some extent. Therefore, we believe that the introduction of this data set can not only expand the amount of data in our experiment but also better explain that our method can have a better segmentation effect on three-dimensional data such as CT.}

\subsection{Implementation Details}

\textbf{Data Preprocessing:} Motivated by the methodologies presented in nnUNet~\cite{isensee2021nnu} and UNETR++~\cite{shaker2024unetr}, our initial step involved an extensive data preprocessing pipeline. Each CT scan image underwent independent preprocessing, where we employed intensity normalization techniques to map values within the [-1000, 1000] Hounsfield Unit (HU) range to the normalized range of [0, 1]. Subsequently, we applied a patch-based cropping strategy to obtain segments of size 192 × 192 × 64.

\textbf{Evaluation Metrics:} The evaluation of our model's performance is grounded in two essential metrics, namely the 95\% Hausdorff Distance (HD95)~\cite{Taha_2015}, Intersection over Union (IoU)
and the Dice Similarity Score (DSC)~\cite{Zou_2004}.

{HD95 is commonly used as a boundary-based metric to measure the 95th percentile of the distance between the boundary predicted by volume partitioning and the ground truth voxel.} It is defined as follows:

\begin{equation}
    HD_{95}\left(Y, P\right)=max\left({d}_{YP}+{d}_{PY}\right)
\end{equation}

Where, ${d}_{YP}$ is the maximum 95th percentile distance between the predicted voxel and the ground truth, and ${d}_{PY}$ is the maximum 95th percentile distance between the ground truth and the predicted voxel.




{DSC is a commonly used evaluation indicator, mainly used to measure the similarity between two sample sets. Its value is between 0 and 1. The larger the value, the higher the similarity between the two sample sets. In this task, it is used to measure the overlap between the label and the model output result. The higher the score, the better the model performance. It is formally defined as follows:}

\begin{equation}
    DSC\left(Y, P\right)=2 \times \frac{\left|Y\cap{P}\right|}{\left|Y\right|\cup{\left|P\right|}}=2 \times \frac{Y\cdot{P}}{Y^2+P^2}
\end{equation}


{Among them, $Y$ and $P$ represent the set of true labels and the predicted results output by the model respectively.}

\textbf{Implementation Details:} Our method was realized through the utilization of PyTorch version 1.10.1, in conjunction with the MONAI library~\cite{cardoso2022monai}. 
The training of our models was executed on two Nvidia GPUs, specifically the 3090Ti 24GB variant, thus optimizing computational efficiency. Notably, the input data was 3D with dimensions set at 192 × 192 × 64. Our training regimen encompassed the utilization of the Adam optimizer, a learning rate of $0.01$, a weight decay parameter set at $3 \times 10^{-5}$, and an expansive training horizon comprising $1000$ epochs.

\begin{figure*}[!h]
    \centering
    \includegraphics[width=0.95\textwidth]{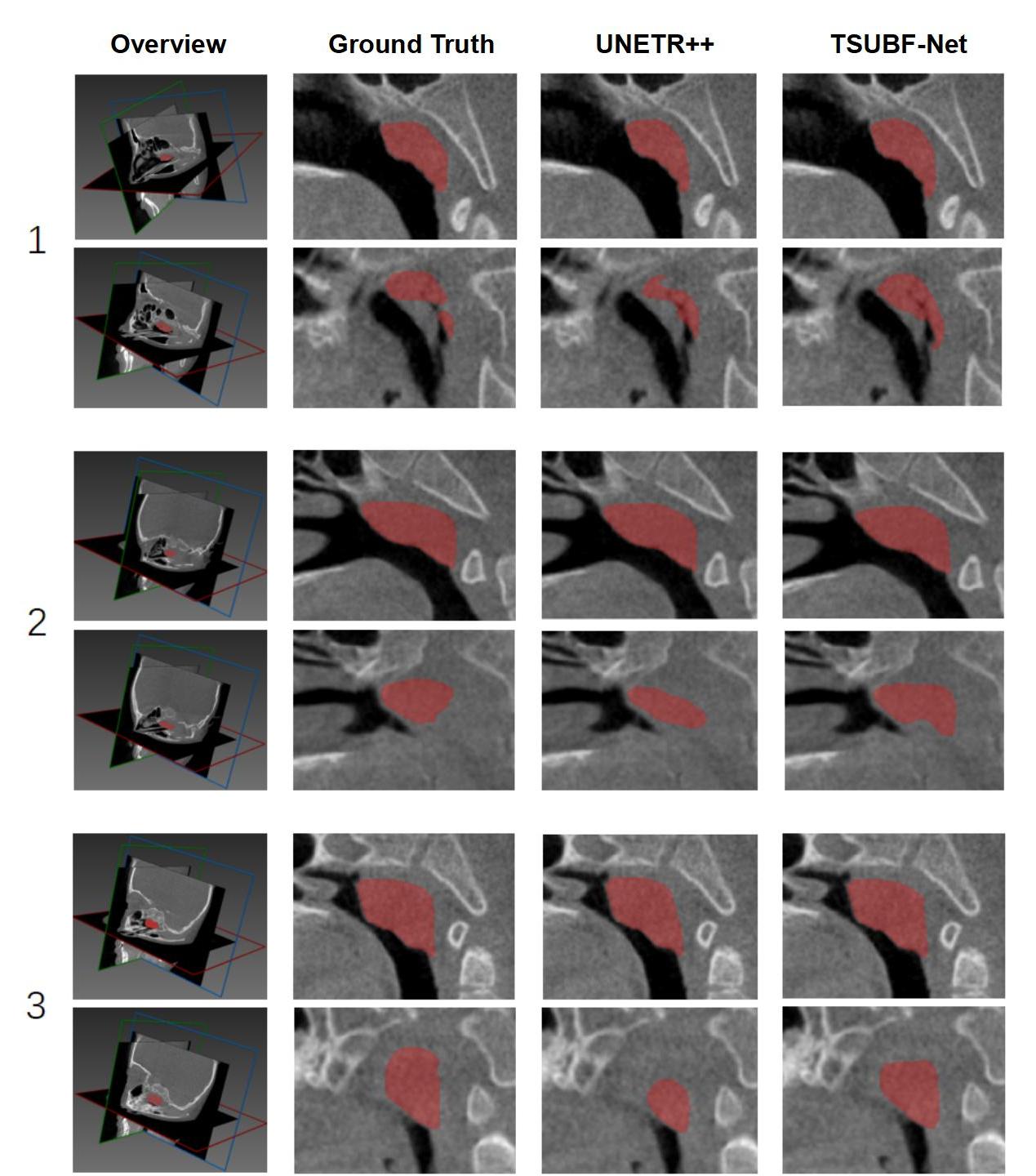}
    \caption{{Qualitative comparison on adenoid hypertrophy segmentation dataset: In the example, we show three segmentation cases as a demonstration. The numbers on the left side of the image indicate the results under the same input data. In the first column, the positions of 2D slice images in the sagittal, coronal, and horizontal planes in CT are shown respectively; the second column shows the display results of the enlarged Ground Truth focus point in the sagittal plane corresponding to CT; in the third column, it is UNETR++ in the corresponding sagittal 2D slice segmentation result image; in the fourth column, it is TSUBF-Net in the corresponding sagittal 2D slice segmentation result image.}}

    \label{fig:experiment}
\end{figure*}

\begin{table}[]
\centering
\renewcommand{\arraystretch}{1.3}
\caption{3D segmentation results on the AHSD Dataset.}
\label{tab:ahsd}
\begin{tabular}{p{0.2\textwidth}p{0.2\textwidth}p{0.2\textwidth}c}
\hline
\textbf{Method}   & \textbf{HD95}    & \textbf{{IoU}}  & \textbf{DSC}      \\ \hline
    U-Net~\cite{ronneberger2015u} & 59.17 & {67.99} & 80.95 \\
    nnUNet~\cite{isensee2021nnu} & 8.16 & {83.71} &  91.13 \\
    UNETR~\cite{hatamizadeh2022unetr} & 31.76 & {72.03} & 83.74 \\
    nnFormer~\cite{zhou2021nnformer} & 13.53 & {78.14} & 88.06 \\
    Swin-UNETR~\cite{hatamizadeh2021swin} & 18.44 & {78.00} & 87.64 \\
    {AgileFormer}~\cite{qiu2024agileformerspatiallyagiletransformer} & {56.32} & {68.08} & {80.98} \\
    UNETR++~\cite{shaker2024unetr} & 10.13 & {79.00} & 88.27 \\
    \hline
    TSUBF-Net & \textbf{7.03} & \textbf{{85.63}} & \textbf{92.26} \\
    \hline
\end{tabular}
\end{table}

\subsection{Results} 
\subsubsection{Adenoid Hypertrophy Segmentation Dataset}
The comparative results for the AHSD are presented in Table~\ref{tab:ahsd}. Notably, performance metrics are reported exclusively based on single-model accuracy, with no utilization of pre-training, model ensembling, or supplementary data. For purely Convolutional Neural Network (CNN)-based methods, the U-Net exhibits a commendable DSC score of 80.95\% but a poor HD95 score. Among the existing hybrid-transformer-CNN-based methods, UNETR and Swin-UNETR deliver promising DSC scores of 83.74\% and 87.64\%, respectively. Notably, our proposed TSUBF-Net surpass the UNETR++ model in terms of the highest DSC value, registering an impressive DSC of 92.26\%. Furthermore, TSUBF-Net attains superior results, as evidenced by its lowest HD95 metric. These results collectively underscore the remarkable performance of TSUBF-Net on this dataset.

{Interestingly, AgileFormer, which has relatively good performance on other datasets, has lower performance in AHSD. This is because AgileFormer is not a true 3D segmentation model.AgileFormer processes 3D data by performing slicing operations along the z-axis and thus as model inputs, which prevents the model from spatially learning in the z-axis direction. Especially in AHSD tasks where geometric boundaries are missing and segmentation results need to be obtained by relative spatial position and continuity of the segmentation plane, AgileFormer has difficulty in learning the correct segmentation method.}

In Fig.~\ref{fig:experiment}, we present a qualitative juxtaposition between the baseline model and the proposed TSUBF-Net in the context of AHSD. Our illustration employs three distinct segmentation cases to provide a comprehensive demonstration, which demonstrates the superiority of our approaches.

{Our proposed approach exhibits substantial performance enhancements across various facets of adenoid hypertrophy. Our visualization comparisons underscore the impressive segmentation capabilities of the proposed methods. Notably, when confronted with indistinct regions or blurred edges in the initial predictions, our approach adeptly leverages inter-layer relationships to yield smoother and more rationalized segmentation predictions.}

\begin{figure*}[!h]
    \centering
    \includegraphics[width=0.95\textwidth, trim=10 10 10 10]{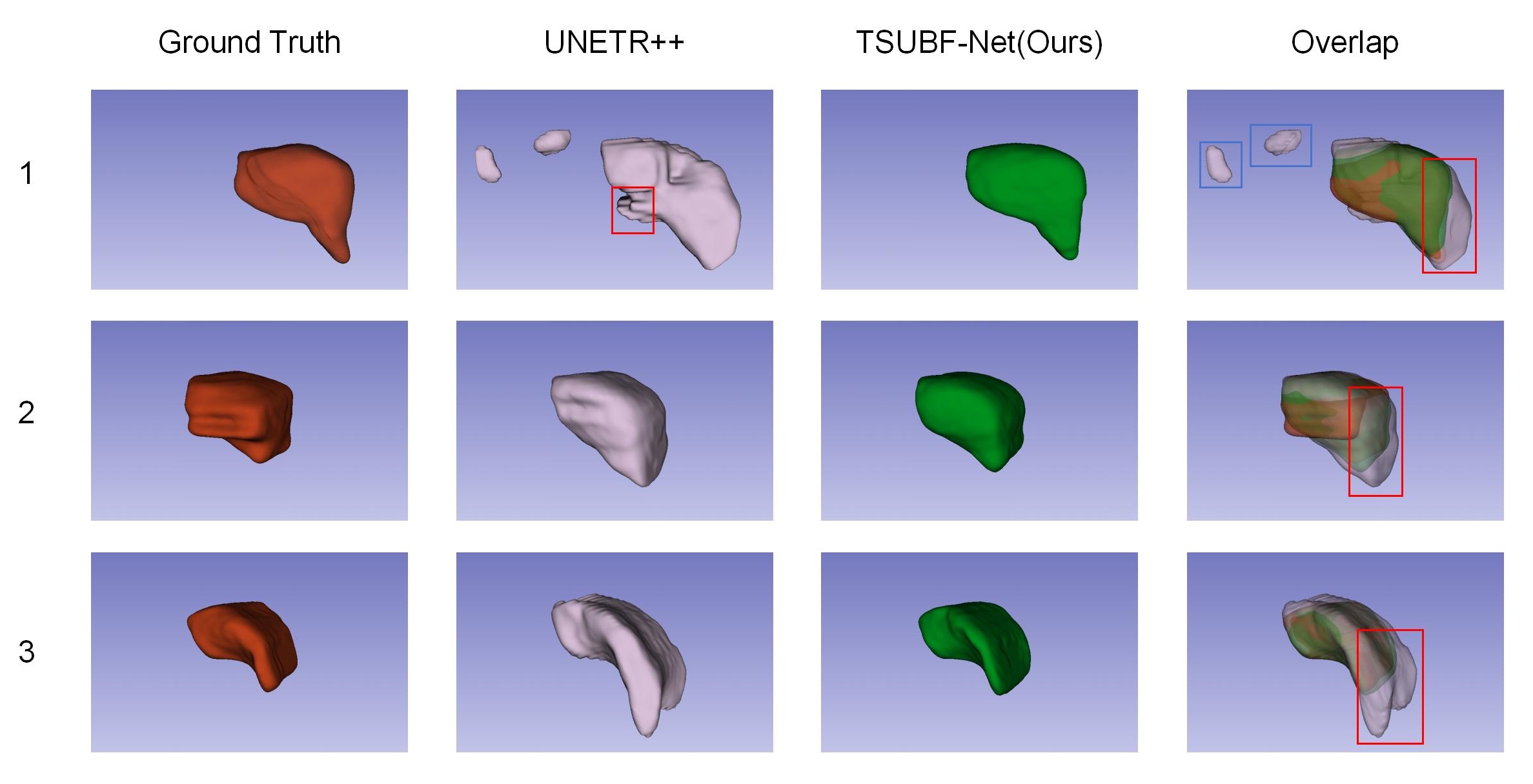}
    \caption{{The vertical axis represents Ground Truth, UNETR++, and TSUBF-Net, The last part is the visualization result after adjusting the transparency of the three components and overlaying them at the same position. The horizontal axis represents three different samples.}}

    \label{fig:compare}
\end{figure*}

{In order to more clearly see the comparison between the result of our segmentation and the result of GroundTruth and UNETR++, we have carried out a three-dimensional visualization of the result, as shown in Fig.~\ref{fig:compare}. Red is the Ground Truth, white is the segmentation result of UNETR++, and green is the segmentation result of TSUBF-Net. In order to see the intersection of the three, we adjust the transparency of the three to see the effect of the final split. It is obvious that there is still a big gap between UNETR++ and our method in dealing with boundary problems, and the lower the part, the more prominent its boundary. From the right side, we can see that our segmentation results are similar to the Ground Truth, both of which have relatively smooth surfaces. However, the segmentation results of UNETR++ in this part are not smooth enough, and there are many small and uneven places. As can be seen from the first sample, the red box indicates that there is still a lot of room for progress on the front and back boundaries of UNETR++. In addition, as shown in the blue box, there are some noise points that are divided into other places. The reason may be that for images with weak boundaries, the processing of UNETR++ is not as good as our proposed method.}

\subsubsection{Automatic Cardiac Diagnosis Challenge}
{Since there are no other publicly available adenoid hypertrophy datasets, we supplemented our experiments with the ACDC dataset to further investigate the performance of TSUBF-Net for the smoothness of segmentation results. We use the DSC metric to evaluate the results of our method on the ACDC dataset.} The comparative analysis results are presented in Table~\ref{tab:acdc}. The average DSC for UNETR++ and AgileFormer stands at 92.83\% and 92.55\%, respectively. And the average DSC of TSUBF-Net reaches 92.68\%, which is higher than that of AgileFormer and other models, and only slightly lower than that of UNETR++. It is particularly noteworthy that TSUBF-Net excels in segmenting the myocardium (Myo) and left ventricle (LV) with accuracy scores of 90.67\% and 96.27\%. Overall, TSUBF-Net emerges as one of the top-performing models in this domain.

\begin{table}[]
\centering
\renewcommand{\arraystretch}{1.3} 
\caption{{State-of-the-art comparison on ACDC. We report the performance on the right ventricle (RV), left ventricle (LV), and myocardium (MYO) along with mean results using the DSC metric.}}
\label{tab:acdc}
\begin{tabular}{c@{\hspace{35pt}}c@{\hspace{20pt}}c@{\hspace{20pt}}c@{\hspace{35pt}}c}
\hline
\textbf{Methods} & \textbf{RV} & \textbf{Myo} & \textbf{LV} & \textbf{Average} \\
\hline
        UNETR~\cite{hatamizadeh2022unetr} & 85.29 & 86.52 & 94.02 & 86.61 \\
        nnFormer~\cite{zhou2021nnformer} & 90.94 & 89.58 & 95.65 & 92.06 \\
        Swin-UNet~\cite{2021Swin} & 88.55 & 85.62 & 95.83 & 90.00 \\
        TransUNet~\cite{chen2021transunet} & 88.86 & 84.53 & 95.73 & 89.71 \\
        MISSFormer~\cite{huang2022missformer} & 86.36 & 85.75 & 91.59 & 87.90 \\
        nnUNet~\cite{isensee2021nnu} & 90.96 & 90.34 & 95.92 & 92.41 \\  
        MedNeXt~\cite{roy2024mednexttransformerdrivenscalingconvnets} & 89.37 & 89.55 & 95.37 & 91.43 \\        
        {AgileFormer}~\cite{qiu2024agileformerspatiallyagiletransformer} & {91.05} & {90.40} & {96.19} & {92.55} \\
        UNETR++~\cite{shaker2024unetr} & \textbf{92.26} & \ 90.61 & 96.00 & \textbf{92.83} \\
        \hline
        TSUBF-Net & 91.13 & \textbf{90.67} & \textbf{96.24} & 92.68 \\
        \hline
\end{tabular}
\end{table}

\begin{figure*}[h]
    \centering
    \includegraphics[width=1\textwidth]{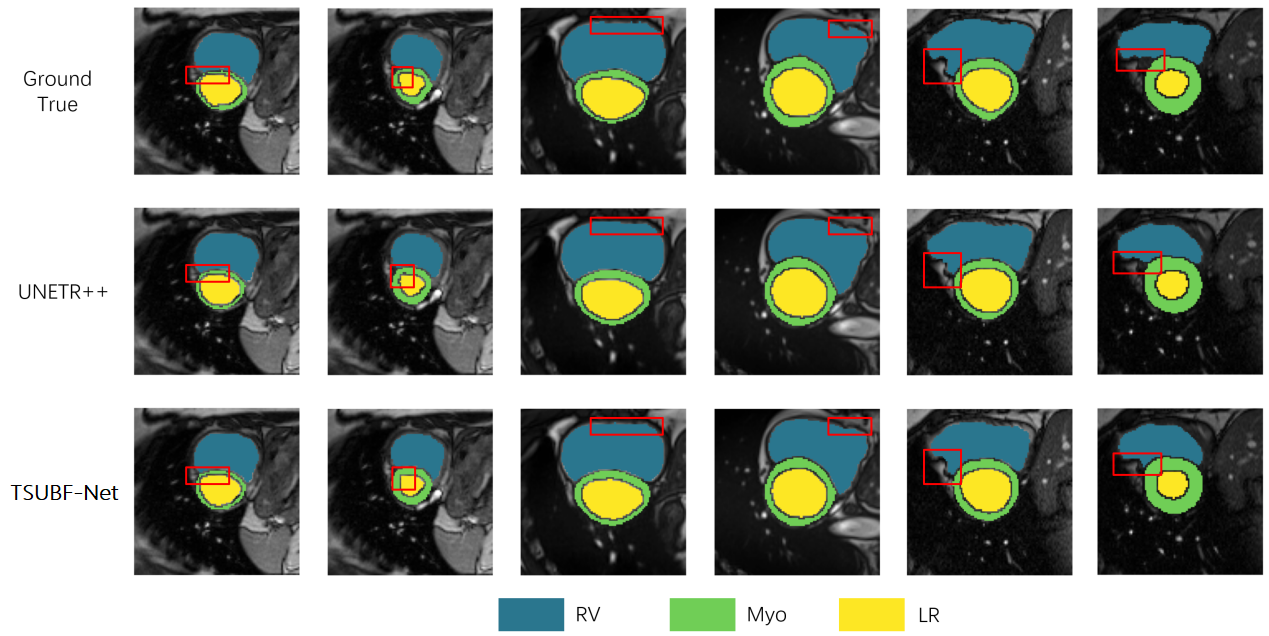}
    \caption{Qualitative comparison on the ACDC dataset. The red box makes it easy to see the detailed comparison of our model with the real values and UNETR++ in boundary processing.}
    \label{fig:acdc}
\end{figure*}

In Fig.~\ref{fig:acdc}, we present a qualitative comparison involving the application of our proposed TSUBF-Net to the ACDC dataset. A focused examination of the outcomes within the delineated red box reveals noteworthy insights. Although our proposed model may not exhibit exceptional performance when applied to publicly accessible organ segmentation datasets, it distinguishes itself through its remarkable capacity to generate results characterized by a superior degree of smoothness.

Furthermore, our proposed model demonstrates a particular adeptness in producing smoother predictions, particularly when the Ground Truth labels exhibit reduced inherent smoothness. Notably, UNETR++ delivers results of a comparable quality, yet our proposed model demonstrates a subtle, but discernible, edge in terms of smoothness, especially in select intricate case details. These findings emphasize the nuanced strengths of TSUBF-Net in addressing specific intricacies within the segmentation task.

{
\subsubsection{MSD-Lung Dataset}
We additionally used the MSD-Lung dataset as a supplement to our experiments to further investigate the segmentation performance of TSUBF-Net. The results of the comparative analyses are shown in Table ~\ref{tab:lung}. The average DSC for UNETR++ and MedNeXt stands at 80.68\% and 80.14\%, respectively. Remarkably, TSUBF-Net consistently outperforms conventional 3D segmentation models in medical imaging, achieving an average DSC of 83.69\%, outperforming UNETR++ and MedNeXt by 3.73\% and 4.4\%, respectively. Overall, TSUBF-Net emerges as one of the top-performing models in this domain, slightly outperforming UNETR++ in terms of average accuracy (Average).
}

\begin{table}[]
\centering
\renewcommand{\arraystretch}{1.3} 
\caption{{3D segmentation results on the MSD-Lung Dataset.}}
\label{tab:lung}
\begin{tabular}{c@{\hspace{20pt}}c@{\hspace{20pt}}c@{\hspace{20pt}}c} 
\hline
\textbf{{Methods}} & \textbf{{HD95}} & \textbf{{IoU}} & \textbf{{DSC}} \\
\hline
        {UNETR~\cite{hatamizadeh2022unetr}} & {23.84} & {57.84} & {73.29} \\
        {nnFormer~\cite{zhou2021nnformer}} & {16.25} & {63.87} & {77.95} \\
        {Swin-UNETR~\cite{hatamizadeh2021swin}} & {28.74} & {60.71} & {75.55} \\
        {nnUNet~\cite{isensee2021nnu}} & {28.52} & {59.12} & {74.31} \\
        {AgileFormer}~\cite{qiu2024agileformerspatiallyagiletransformer} & {3.04} & {65.01} & {78.80} \\
        {MedNeXt~\cite{roy2024mednexttransformerdrivenscalingconvnets}} & {2.85} & {66.86} & {80.14} \\
        {UNETR++~\cite{shaker2024unetr}} & {2.79} & {67.62} & {80.68} \\
\hline
        {TSUBF-Net} & \textbf{{2.37}} & \textbf{{71.95}} & \textbf{{83.69}} \\
\hline
\end{tabular}
\end{table}

\subsubsection{Results of Ablation Experiment}

In Table~\ref{tab:baseline}, we present a comprehensive evaluation of the impact of our proposed methods on the baseline UNETR++ framework for the adenoid hypertrophy segmentation task. The efficacy of the models is quantitatively assessed using the Dice Similarity Coefficient (DSC), which serves as a robust indicator of segmentation accuracy. It is important to note that all reported results are based on the evaluation of individual model accuracy.

\begin{table}
\renewcommand{\arraystretch}{1.3} 
    \caption{{Ablation study on our proposed methods. Results are shown in terms of the segmentation algorithm model Floating Point Operations Per Second (FLOPs) and performance (HD95, IoU, and DSC). $\lambda$ is a coefficient with respect to $L_\text{sobel}$. For a fair comparison, all the results use the same input size and preprocessing, and it can be seen the impact of different modules on each evaluation metric.}}
    \label{tab:baseline}
    \centering
    \begin{tabular}{cccc|ccccc} 
    \hline
        \textbf{Baseline} & \textbf{TSP} & \textbf{BSCF} & \textbf{$\lambda$} & \textbf{FLOPs(G)} & \textbf{HD95} & \textbf{{IoU}} & \textbf{DSC} & \textbf{Enhancement} \\
        \hline
        \textbf{\checkmark} & / & / & 0 & 134.07 & 10.17 & {79.32} & 88.22 & 0\% \\
        \textbf{\checkmark} & \textbf{\checkmark} & / & 0 & \textbf{92.2} & 12.53 & {78.82} & 87.78 & -0.44\% \\
        \textbf{\checkmark} & / & \textbf{\checkmark} & 0 & 196.34 & 8.23 & {83.33} & 90.91 & 2.69\% \\
        \textbf{\checkmark} & \textbf{\checkmark} & \textbf{\checkmark} & 0 & 154.47 & 7.89 & {84.31} & 91.49 & 3.27\% \\
        {\textbf{\checkmark}} & {\textbf{\checkmark}} & {\textbf{\checkmark}} & {1.0} & {154.47} & {7.78} & {84.63} & {91.68} & {3.46\%} \\
        {\textbf{\checkmark}} & {\textbf{\checkmark}} & {\textbf{\checkmark}} & {0.5} & {154.47} & {7.73} & {84.68} & {91.71} & {3.49\%} \\
        \textbf{\checkmark} & \textbf{\checkmark} & \textbf{\checkmark} & 0.1 & 154.47 & \textbf{7.03} & \textbf{{85.63}} & \textbf{92.26} & \textbf{4.57\%} \\
    \hline
    \end{tabular}
\end{table}

Our evaluation reveals that the enhancements within the modules lead to improvements in the context of the adenoid segmentation dataset. Specifically, the TSP module has a slight decrease in accuracy due to a lower parameter than the UNETR++'s EPA module. However, the incorporation of the BSCF module yields more substantial enhancements, resulting in a notable improvement of 2.69\% when compared to the baseline. Furthermore, the combination of the two modules culminates in a significant DSC enhancement of 3.27\%. The most noteworthy augmentation is observed when the new loss function is employed, leading to the highest recorded DSC value of 0.9226, signifying an impressive improvement of 4.57\% compared to the baseline. These findings underscore the effectiveness of the proposed modifications in enhancing the segmentation accuracy of TSUBF-Net for the adenoid hypertrophy task.{We also use different $\lambda$ values to replace the hyperparameters in the loss function, and the $\lambda$ values mainly affect the loss part. In our experiments, we found that the best score was achieved at 0.1, with a 4.57\% Enhancement. }

{The complexity of the model has a direct impact on the training and inference of the model and is positively correlated. We calculated the time spent in training 1000 rounds of the model under different FLOPs(G) and the time spent in reasoning a sample. When FLOPs (G) =92.2, it takes 96 hours for model training and 7 minutes for inference of a single sample. When FLOPs (G) =134.07, it takes 99 hours for model training and 9 minutes for inference of a single sample. When FLOPs (G) =154.47, the model training takes 102 hours and the inference for a single sample takes 10 minutes. Although our method is compared with UNETR++, the model training and reasoning time are slightly longer than theirs, but our accuracy and other indicators are completely beyond their indicators. }



\subsection{Discussion}

{
In this section, we mainly discuss our experiments in three datasets and the ablation experiments. From the perspective of our adenoid dataset (AHSD), our results in the three evaluation indicators are far better than other models, and the scores of HD95, IoU and DSC are 7.03, 85.63, and 92.26 respectively. There is a significant improvement in the index of IoU. Although only a small improvement over nnUNet, it is limited by its complex model structure at the expense of computational efficiency and time complexity. In addition, compared with UNETR++, our TSP module can greatly reduce the computational complexity, and BSCF module effectively solves the boundary merging problem based on segmentation data. However, in tasks involving clear boundaries, this design may introduce noise and reduce accuracy. Despite these challenges, TSUBF-Net's design principles emphasize interlayer relationships within the loss function, significantly improving segmentation smoothness, so that according to our two innovative modules and the loss function, the model ultimately supports a good balance between accuracy and smoothness in performance.}

{In experiments with the two publicly available datasets we use, our methods also outperform current algorithms. In the ACDC dataset, the DSC values of Myo and LV segments are at a leading level. In the MSD-Lung dataset, we far exceed the current comparison model in the three indexes, reaching HD95:2.37, IoU:71.95, and DSC:83.69. We find that the reason why our TSUBF-Net does not reach the best level in all indicators on ACDC is because this dataset is MRI data and the data quality and imaging mode are different from CT. Another reason is that we are overly focused on boundary smoothness when partitioning tasks, and not all tasks require smooth edges like hypertrophy segmentation of adenoids. }

{For the ablation experiment, we conducted ablation for TSP and BSCF modules. In particular, we also conducted ablation experiments for the hyperparameters in the loss function to verify the influence of hyperparameters on each result. In Table 5, comparing the first and second lines, we find that after replacing the Baseline Model's EPA module with our TSP module, the computational complexity is greatly reduced by 41.87 FLOPs (G), but the accuracy is also slightly decreased by 0.44\%. We think that our TSP module can reduce a lot of complexity at the expense of a little precision. In addition, the addition of our proposed BSCF module can greatly improve the accuracy of the model, although it also increases by 62.27 FLOPs (G). In summary, the two modules proposed by us can balance the computational complexity and precision indexes well. For the ablation part of the loss function, we set a total of three different values for the ablation experiment, namely $\lambda$ = 1.0, 0.5, 0.1, and 0. According to the results, when the $\lambda$ is set to 1.0, the three evaluation indicators all get the best results. However, we believe that there is still some room for improvement. According to the method we designed, the loss is only used as a regular term to adjust the edge, making the segmentation result smoother. However, as mentioned above, not all datasets have this requirement, so we believe that this $\lambda$ value can be learned. After some feature extraction of data and labels in the model, $\lambda$ is obtained.}

\section{Conclusion}

{
In this paper, we propose a framework named TSUBF-Net, designed for the 3D medical segmentation of adenoids. To address the challenges of fuzzy boundaries, we design three innovative components. The Trans-Spatial Perception (TSP) module is introduced to effectively encode continuous spatial and independent channel features, which enhance the model’s ability to process features within the 3D medical data. Additionally, we introduce the Bi-direction Sample Collaborated Fusion (BSCF) module optimized for the nuances of 3D medical segmentation tasks. Moreover, the Sobel loss term is introduced to optimize the segmentation trauma smoothness. Notably, these enhancements enable TSUBF-Net to achieve superior segmentation results on the adenoid segmentation dataset, while concurrently enhancing the smoothness of segmentation outcomes. Our TSUBF-Net also achieved better performance than the most advanced methods on two different datasets, ACDC(CT) and MSD-Lung(MRI), which solidifies the usefulness of TSUBF-Net in solving more organ segmentation challenges and has the potential to solve other organ 3D segmentation challenges. However, there are still limitations in the current research. Our model excessively pursues the smoothness of the edge, and there is still much room for improvement in its own segmentation performance and the boundary certainty of the back boundary. In the future, we will continue to explore more methods in the adenoid hypertrophy task, not only the adenoid hypertrophy part, but also hope to add the tonsil, turbinate, epiglottis, and other parts to our segmentation task to build a complete database of sleep apnea disease. }


\bigskip\noindent \textbf{Acknowledgements} This work was supported by the National Natural Science Foundation of China (No. 52275565), NSF of Guangdong province (No. 2022A1515011667), Shenzhen Science and Technology Program (No. JCYJ20200109114244249), and Youth Talent Fund of Guangdong province (No. 2023A1515030292). Acknowledgment of the computing resource from HPC of INSE in Shenzhen university.

\bigskip\noindent \textbf{Author Contributions} \textbf{Rulin Zhou:} Conceptualization, Methodology, Validation, Investigation, Original draft preparation. \textbf{Yingjie Feng:} Conceptualization, Methodology, Validation, Investigation, Original draft preparation. \textbf{Guankun Wang:} Investigation, Methodology, Original draft preparation. \textbf{Xiaopin Zhong:} Supervision, Resources. \textbf{Zongze Wu:} Supervision,Resources. \textbf{Qiang Wu:} Supervision, Resources. \textbf{Xi Zhang:} Conceptualization, Supervision, Review, and editing, Re-sources.

\bigskip\noindent \textbf{Data Availability} In this study, two datasets were mainly used: (a) the Adenoid Hypertrophy Segmentation Dataset (AHSD), which was collected in collaboration with Shenzhen University General Hospital. Due to patient privacy reasons, the data cannot be publicly disclosed. (b) Another dataset for heart segmentation comes from the 2017 Automated Cardiac Diagnosis Challenge (ACDC), which can be found on the official website(https://www.creatis.insa-lyon.fr/Challenge/acdc/index.html )

\section*{Declarations}

\bigskip\noindent \textbf{Conflicts of interest} All the authors declare that there are no interest conflicts that could influence the work presented in this article.

\bigskip\noindent \textbf{Ethical and informed consent for data used} The Adenoid Hypertrophy Segmentation Dataset (AHSD) collected from Shenzhen University General Hospital in this study underwent strict ethical review during the data collection process and was officially approved by the Ethics and Welfare Committee of Shenzhen University General Hospital (Approval Number: KYLL-20230402A). All content in the experiment has obtained informed consent from the patient.

\bibliography{sn-bibliography}

\end{document}